 \documentclass[CEJP,DVI]{cej} 
\usepackage{layout}
\usepackage{amsmath}
\usepackage{textcomp}
\usepackage{hyperref}
\usepackage{epsf}

\title{Equation of state at non-zero baryon density based on lattice 
       QCD}

\articletype{Communication\footnote{Presented at CPOD 2011, November 7-11, 2011,
                       Wuhan, China}}

\author{Pasi~Huovinen\inst{1}\email{huovinen@th.physik.uni-frankfurt.de},
        P\'eter~Petreczky\inst{2},
        Christian~Schmidt\inst{3}}

\institute{
     \inst{1} Institut f\"ur Theoretische Physik,
              Johann Wolfgang Goethe-Universit\"at,\\
              60438 Frankfurt am Main, Germany
     \inst{2} Physics Department, Brookhaven National Laboratory,\\ 
         Upton, NY 11973, USA
     \inst{3} Fakult\"at f\"ur Physik, Universit\"at Bielefeld,\\
              33615 Bielefeld, Germany
          }

\abstract{We employ the lattice QCD data on Taylor expansion coefficients to
  extend our previous parametrization of the equation of state to
  finite baryon density. When we take into account lattice spacing and
  quark mass dependence of the hadron masses, the Taylor coefficients
  at low temperature are equal to those of hadron resonance gas.
  Parametrized lattice equation of state can thus be smoothly
  connected to the hadron resonance
  gas equation of state at low temperatures.}


\begin{document}
\maketitle


One of the methods to extend the lattice QCD calculations to non-zero
chemical potential is Taylor expansion. In that approach pressure is
Taylor expanded in chemical potentials, and the Taylor coefficients
are calculated on the lattice at zero chemical potential. In this
contribution we use the results of the most comprehensive lattice QCD
analysis of the Taylor coefficients to date~\cite{Miao:2008,Cheng:2008}
to construct a parametrisation of an equation of state (EoS) for
finite baryon density. As in our earlier parametrisation of the EoS at
zero chemical potential~\cite{Huovinen:2009}, we require that our
parametrisation matches smoothly to the hadron resonance gas (HRG) at
low temperatures.

\begin{figure}[b]
   \hfill
  \epsfysize=60mm \epsfbox{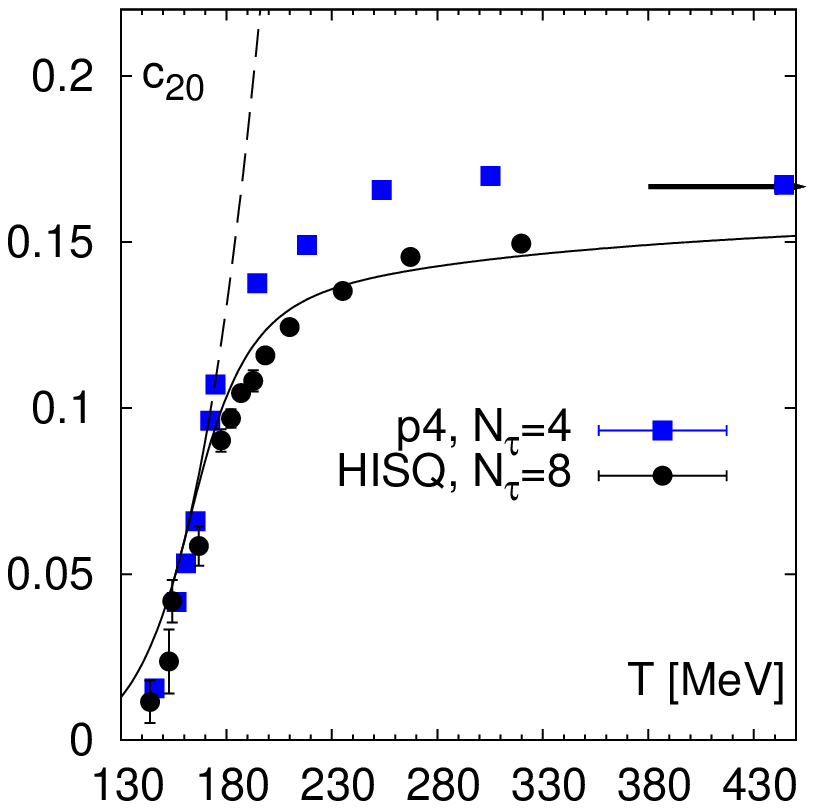}
   \hfill
  \epsfysize=60mm \epsfbox{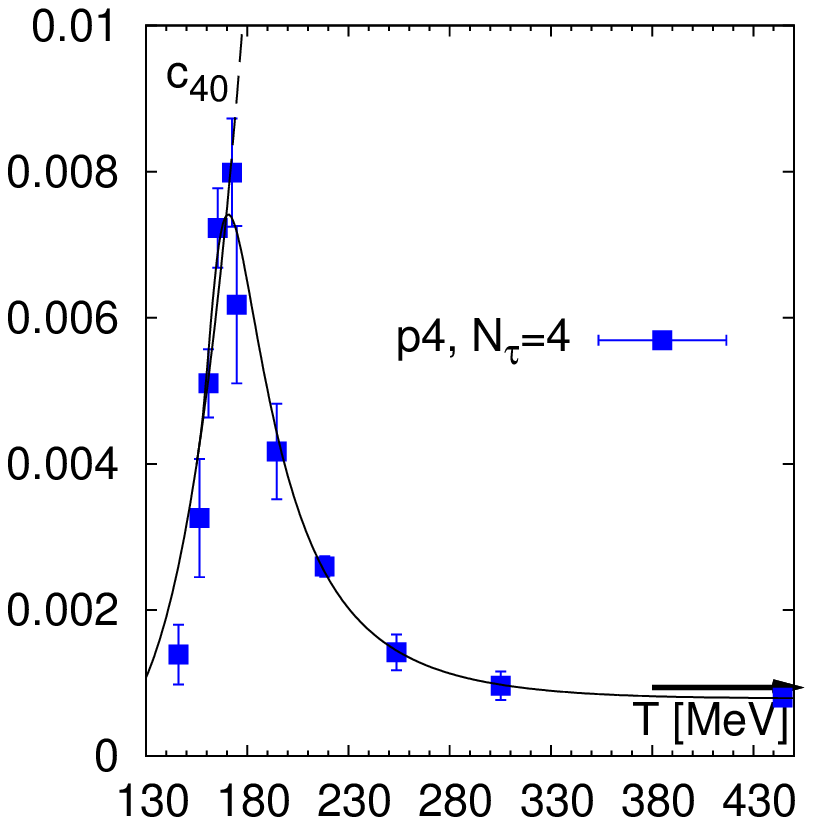}
   \hfill
  \caption{The parametrisation (solid line) and HRG value
    (dashed) of the $c_{20}$ (left) and $c_{40}$ (right) coefficients
    compared with the shifted p4 data (see the text). The recent
    lattice result for $c_{20}$ with the HISQ action~\cite{Bazavov} is also
    shown. The arrows depict the Stefan-Boltzmann values.}
 \label{fig:taylor}
 \end{figure}

Taylor coefficients are derivatives of
pressure, $P$, with respect to baryon and strangeness chemical 
potentials, $\mu_B$ and $\mu_S$, respectively:
\begin{equation}
 c_{ij}(T) = \frac{1}{i!j!}\frac{T^{i+j}}{T^4}\frac{\partial^i}{\partial \mu_B^i}
             \frac{\partial^j}{\partial \mu_S^j}P(T,\mu_B=0,\mu_S=0),
\end{equation}
where $T$ is temperature\footnote{We use natural units where 
$c=\hbar=k_\mathrm{B}=1$ throughout the text.}.
As we discussed in~\cite{QM}, all the coefficients evaluated in
Refs.~\cite{Miao:2008,Cheng:2008} are well below the HRG results.
This discrepancy can largely be explained by the lattice
discretisation effects on hadron masses: When the hadron mass spectrum
is modified accordingly (for details see~\cite{Huovinen:2010}), the
HRG model reproduces the lattice data, see Fig.~1 of Ref.~\cite{QM}.
Interestingly this change can be accounted for by shifting the
modified HRG result of purely baryonic coefficients towards lower
temperature by 30 MeV. The situation is similar for other Taylor
expansion coefficients~\cite{Huovinen:2010}, although the strange
coefficients might favour slightly smaller shift. Based on this
finding and because the lattice data agree so well with the modified
HRG we suggest that cutoff effects can be accounted for by shifting
the lattice data by 30 MeV.  We show the effect of such a shift in the
left panel of Fig.~\ref{fig:taylor}, where we plot the HRG curve with
physical masses (dashed line) and compare it with the lattice data,
where all the points below 206 MeV temperature are shifted by 30 MeV,
and the 209 MeV point by 15 MeV. For further confirmation of this
procedure we also plot the recent HISQ result of
$c_{20}$~\cite{Bazavov} in Fig.~\ref{fig:taylor} (right): At low
temperatures the shifted p4 data agree with the HISQ data.

\begin{figure}[t]
  \epsfysize=60mm \epsfbox{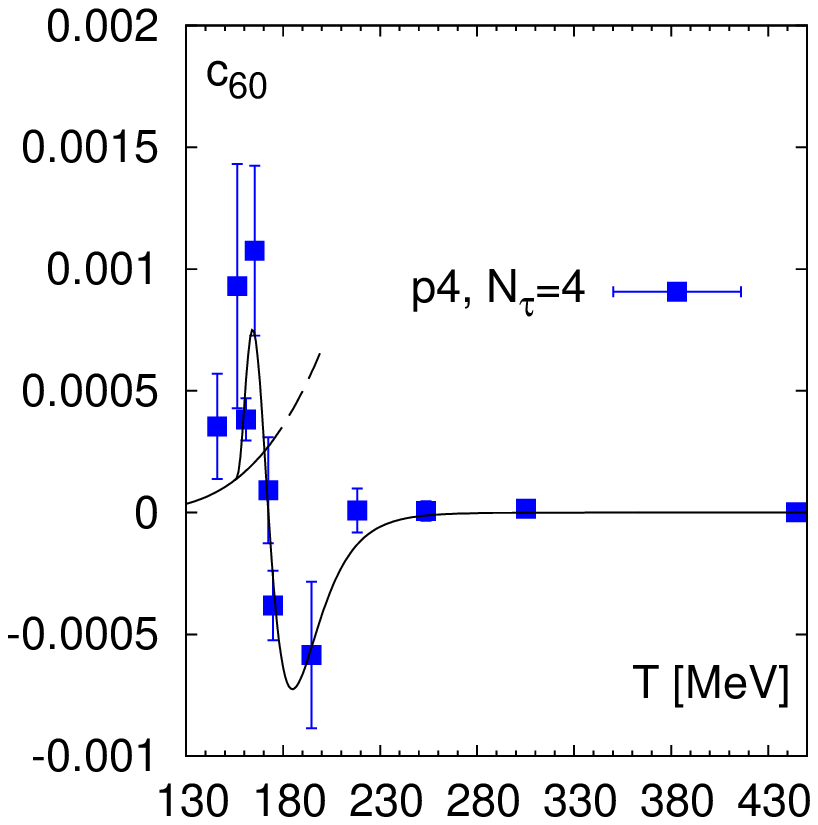}
   \hfill
  \epsfysize=60mm \epsfbox{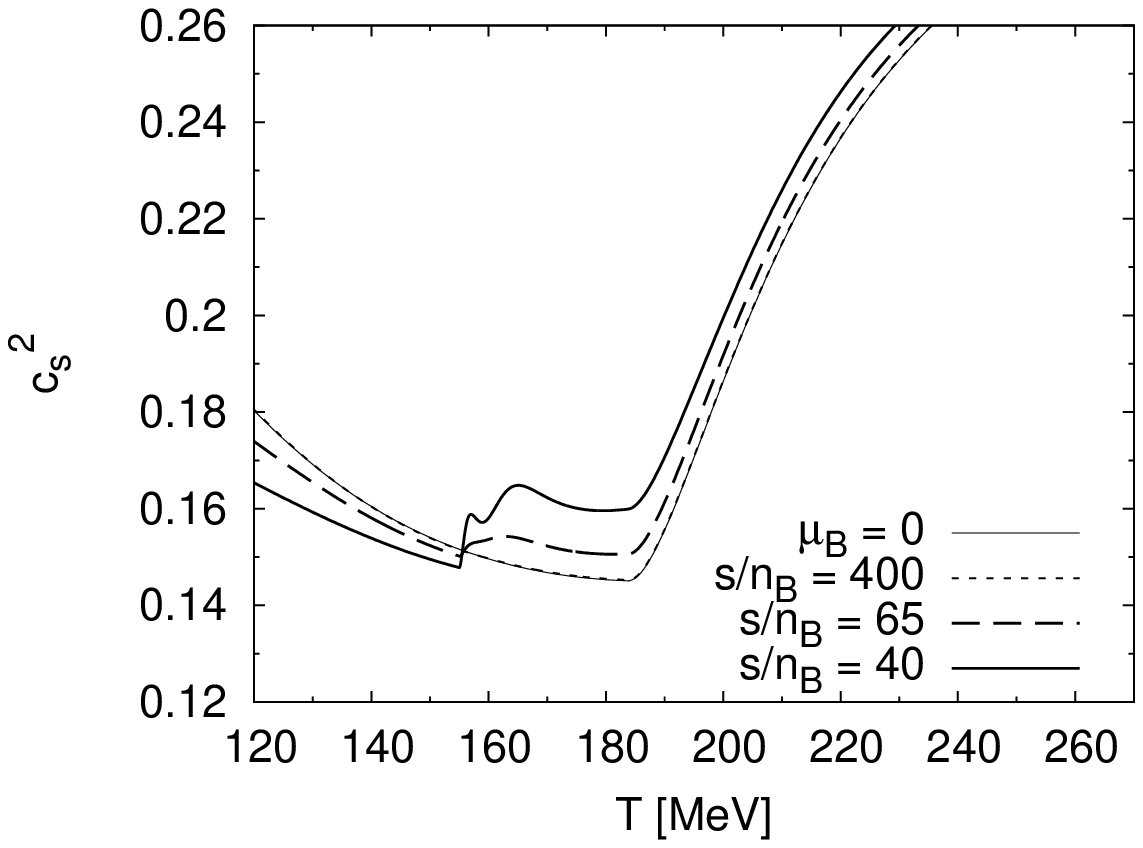}
  \caption{(Left) The parametrisation (solid line) and HRG value
    (dashed) of the $c_{60}$ coefficient compared with the shifted p4
    data (see the text).\\  (Right) The square of the speed of sound 
    $c_s^2$ as a 
    function of temperature on various isentropic curves with constant
    entropy per baryon $s/n_B$ (see the text).}
 \label{fig:eos}
 \end{figure}

We parametrise the shifted data as a function of temperature $T$ 
using an inverse polynomial of four
 ($c_{20}$), five ($c_{11}$ and $c_{02}$), or six (fourth and sixth
 order coefficients) terms:
\begin{equation}
   c_{ij}(T) = \sum_{k=1}^m \frac{a_{kij}}{\hat{T}^{n_{kij}}} + c_{ij}^\mathrm{SB},
\end{equation}
where $c_{ij}^\mathrm{SB}$ is the Stefan-Boltzmann value of the
particular coefficient, $a_{kij}$ are the parameters,
and the powers $n_{kij}$ are required to be
integers between 1 and 42. $\hat{T} = (T-T_s)/R$ with scaling factors
$T_s=0$ and $R=0.15$ GeV for the second order coefficients and
$T_s=0.1$ GeV and $R=0.05$ GeV for all the other coefficients.
We match this parametrisation to the HRG
value at temperature $T_\mathrm{SW}=155$ MeV by requiring that the
Taylor coefficient and its first, second, and third derivatives are
continuous. Since the recent lattice data obtained using HISQ
action~\cite{Bazavov} shows that the second order coefficients
approach their Stefan-Boltzmann limits slowly, we require that their
value is 95\% of their Stefan-Boltzmann value at 800 MeV
temperature. These constraints fix four (or five) of the parameters
$a_{kij}$. The remaining parameters, are fixed by a $\chi^2$ fit to
the lattice data. As an example we show the parametrised $c_{20}$, $c_{40}$,
and $c_{60}$ in Figs.~\ref{fig:taylor} and~\ref{fig:eos}.

Once the coefficients are known, pressure can be written as
\begin{equation}
\frac{P}{T^4} = \sum_{ij} c_{ij}(T) \left(\frac{\mu_B}{T}\right)^i
                                  \left(\frac{\mu_S}{T}\right)^j,
 \label{PT4}
\end{equation}
and all the other thermodynamical quantities can be obtained from
Eq.(\ref{PT4}) by using the laws of thermodynamics. This kind of an
expansion breaks down at large chemical potentials. 
However, at baryon densities of
interest here, the contribution from coefficients of particular order
is clearly below ($<20$\%) the lower order contribution. Thus the
approximation of the EoS in terms of fourth- and sixth order expansion
looks reasonable. As pressure at $\mu_B=0$, i.e.\ the coefficient
$c_{00}$, we use our earlier parametrisation
$s95p$-v1~\cite{Huovinen:2009}. We describe the EoS in the right panel
of Fig.~\ref{fig:eos} by showing the square of the speed of sound 
$c_s^2 = \partial P/\partial\epsilon|_{s/n_B}$~\cite{Teaney}
on various
isentropic curves with constant entropy per baryon $s/n_B$~\cite{Schmidt}.
The curves at $s/n_B=400$, 65, and 40 are relevant at collision
energies $\sqrt{s_\mathrm{NN}}=200$, 39 and 17 GeV, respectively. At
$s/n_B=400$ (dotted line), the EoS is basically identical to the EoS
at $\mu_B = 0$ (thin solid line).  At larger baryon densities the
effect of finite baryon density is no longer negligible. The larger
the density, the stiffer the EoS above, and softer below the
transition temperature.

Furthermore, additional structure begins to appear around the
transition temperature with increasing density. We expect this
structure to be an artifact of our fitting procedure: Our fit is too
good and elevates errors to features leading to additional ripples in
the speed of sound. Another interesting feature in the equation of
state is the rapid change of the speed of sound around
$T_\mathrm{sw}=155$ MeV and another change around $T\approx 185$
MeV. The latter has its origin in our baseline $\mu_B=0$ EoS. It follows
hadron resonance gas up to $T = 184$ MeV temperature causing a change
in the speed of sound at that temperature. Nevertheless, when pressure
is plotted as a function of energy density, these structures are
hardly visible. Therefore we do not expect them to affect the buildup
of flow and the evolution of the system, and consider our
parametrisation a reasonable first attempt.

We have studied the effect of the EoS on flow by calculating elliptic
flow in Pb+Pb collision at the full SPS collision energy
($\sqrt{s_\mathrm{NN}}=17$ GeV). Our results are similar to those we
have shown earlier~\cite{QM}: Even if in an ideal fluid calculation at
RHIC energy proton $v_2(p_T)$ is sensitive to the order of phase
transition~\cite{Huovinen:2005}, at SPS energy both proton and pion
$v_2(p_T)$ are insensitive to it.

To summarise, we have shown that a temperature shift of 30 MeV is a
good approximation of the discretisation effects in the lattice QCD
data obtained using p4 action. We have constructed an equation of
state for finite baryon densities based on hadron resonance gas and
lattice QCD data. At the full SPS energy ($\sqrt{s_\mathrm{NN}} = 17$
GeV) the $p_T$-differential elliptic flow is almost insensitive to the
equation of state. This is bad news for the experimental search of the
critical point, since a change from a first order phase transition to
a smooth crossover does not cause an observable change in the flow.

{\bf Acknowledgements:}
This work was supported by BMBF under contract no.\ 06FY9092, and by
the U.S. Department of Energy under contract DE-AC02-98CH1086.


\end{document}